\documentclass[aps,pre,twocolumn,showpacs,superscriptaddress]{revtex4}
\usepackage{graphicx}
\usepackage{amsmath}
\bibstyle{apsrev.bib}

\def\ARTICLETYPE{{paper}}
\def\WIDTHB{0.42\textwidth}  
\def\WIDTHA{0.48\textwidth}  
\newcommand{\beq}{\begin{equation}}
\newcommand{\eeq}{\end{equation}}

\begin{document}

\title{Scale free networks from a Hamiltonian dynamics}
\author{M. Baiesi}
\affiliation{INFM-Dipartimento di Fisica, Universit\`a di Padova, I-35131 Padova, Italy.}
\author{S. S. Manna}
\affiliation{INFM-Dipartimento di Fisica, Universit\`a di Padova, I-35131 Padova, Italy.} 
\affiliation {Satyendra Nath Bose National Centre for Basic Sciences Block-JD, Sector-III, Salt Lake, Kolkata 700098, India}
\date{\today}

\begin{abstract}
Contrary to many recent models of growing networks,
we present a model with fixed number of nodes and links,
where it is introduced a dynamics favoring the formation of links 
between nodes with degree of connectivity as different as possible.
By applying a local rewiring move,  the network reaches
equilibrium states assuming broad degree distributions, which have a
power law form in an intermediate range of the parameters used.
Interestingly, in the same range we find non-trivial hierarchical clustering.
\end{abstract}


\maketitle

In their theory of random graphs (RG), Erd\"os and {R\'enyi} showed
that these graphs, composed of $N$ vertices (or nodes), 
connected probabilistically
by a set of edges have several interesting properties~\cite {Erdos}. Among them
the most striking one is the slow rate of growth  ($\sim \log N$) of the
diameter of the giant component. This ``small world'' property
is very important in connected networks represented by single
component graphs, since it reflects the efficiency of the network
for transport or communications~\cite {WS}. Over last few years it is becoming
increasingly evident that most real-world networks have indeed small
world properties~\cite{linked,BA},
e.g., electronic communication networks like Internet~\cite{Faloutsos},
World-Wide Web~\cite{web},
social networks of acquaintances~\cite{social} and of
collaborations~\cite{colla}.

On the other hand, some important properties distinguish real-world networks  
from RG, motivating the rapid growth of interest in this field.
Many real-world networks have broad nodal degree distributions, 
$P(k)$ (the degree of a node is the number $k$ of links 
meeting at that node) often characterized by a power tail, 
$P(k)\sim k^{-\gamma}$, that indicates
a scale free (SF) character of the network (SFN)~\cite {BA}. 
Moreover, in real-world networks one observes a high degree of clustering,
which measures the local correlations among the links of the network 
and implies that neighbors of a node are more likely to be
 neighbors~\cite{WS} (this feature has also been associated to the term small
world~\cite{WS}).
The clustering often scales with the degree of the relative node. This
is connected to a hierarchical organization of the network~\cite{ravasz03},
where clustered blocks connect to form larger units, and etc.

To introduce the correlations between nodes that distinguish SFN's from RG's,
in the last years SFN's have been extensively modeled by growing networks in 
which a preferential attachment (PA) rule shapes the nodes 
degree~\cite{linked,BA},
(i.e., each new node is linked to an old one with a probability 
proportional to the degree of the old node~\cite {barabasi}).
However, biological networks, including food webs~\cite{food}, 
metabolic networks~\cite{metabol_hier,maslov} and
 protein-protein interaction networks~\cite{protein_Uetz,Bara_Ck} 
display the features listed above, although both the PA and the
growing process are debatable in these cases. 
For example, for protein-protein networks they have proposed
both growing network models without
PA~\cite{protein_VFMV} and models with a dominant 
stationary, asymmetric PA~\cite{protein_BLW}.
In food webs, where links represent the prey-predator relations,
the PA and growth are particularly unsuitable to describe the situation.
Thus, in order to achieve a better understanding of the principles 
shaping a part of the real networks, it is worth to spend some efforts 
to discover dynamics leading to non-growing SFN's.
It has been already shown that models with fixed number of nodes 
 do not require linear PA, but SF distributions arise from an algebraic 
PA rule including an exponent which can vary in a wide
range~\cite{manna}.

\begin{figure*}[!t]
\includegraphics[width=0.95\textwidth]{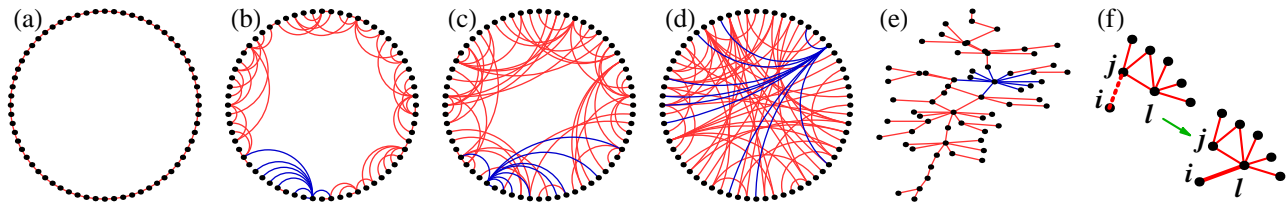}
\caption{
(a)-(d): Example with $N=L=64$, $\bar k=2$ at $\alpha=2$.
Snapshots after $t$ iterations of the LRM (f), 
with (a) $t=0$, (b) $t=4N$, (c) $t=12N$ 
and (d) $t=N^2$.
Blue (darker in gray scale) links meet at the node with the highest degree.
(e): Different layout of (d). (f): LRM described in the text.
\label{fig:ex}}
\end{figure*}

In this \ARTICLETYPE\ we address the particular problem of whether SFN's
can arise from mechanisms excluding both PA and growth.
Recent works \cite{lassig,bauer,burda} proposed theories of
networks at equilibrium. In some of these cases \cite{bauer,burda}
a SFN can be generated simply
by choosing the desired degree distribution to be SF.
On the other hand, different works obtained SFN without plugging in an 
{\it a priori} degree distribution~\cite{caldarelli,valverde,doye,hidden}.
In the spirit of the latter strategy, 
we propose an example of equilibrium network with Hamiltonian that can yield 
hierarchical SFN's.
The energy function depends on the degrees of the nodes and
of their neighbors~\cite{lassig}.
Since it favors connections between nodes with degrees as
different as possible, it leads to networks with disassortative 
mixing~\cite{assort}.
Furthermore, the simulation is implemented by using a local rewiring rule.
Dynamics of this kind appear natural for biological networks, which indeed are
disassortative~\cite{assort,maslov}. In particular, in food webs we expect
each species to find not convenient to interact with similar (and competing)
ones. We notice that in food webs both exponential and SFN's are 
found~\cite{food}, as we have in our model.

We consider a connected network, represented by a single component 
undirected graph,
composed of $N$ nodes connected by $L$ undirected links (edges).
The network topology is uniquely
determined by its adjacency matrix ${\bf c}$, such that $c_{ij}=1$
if nodes $i$ and $j$ are linked, and $0$ otherwise.
The degree of a node $i$ is indicated as $k_i=\sum_{j\neq i} c_{ij}$.
We define an energy associated with a link between
the $i$-th and the $j$-th nodes having the following form: 
 \beq
\epsilon_{ij} = - c_{ij} \left( 1 - \frac{ \min\{ k_i, k_j \} }
{ \max\{ k_i, k_j \} } \right) \,.
\label{eq:eij}
\eeq
      This equation implies that the energy of a link 
      decreases with the difference in nodal degrees at
      the two ends of the link and it contributes no
      energy when the link connects two nodes of same degrees.
The Hamiltonian of a network configuration ${\mathcal G}$ is then
\beq
H(\mathcal G) = \sum_{i<j} \epsilon_{ij} \,.
\label{eq:Hamil}
\eeq

\begin{figure}[!t]
\includegraphics[angle=-90,width=\WIDTHB]{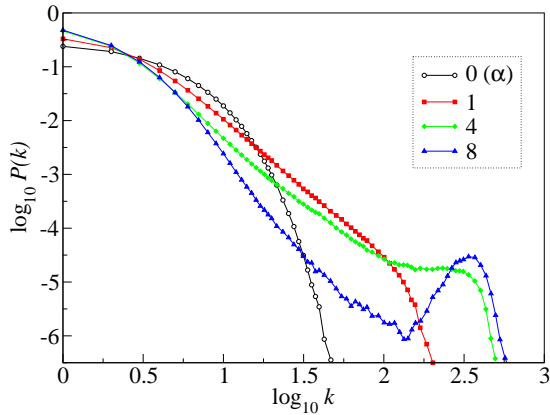}
\caption{Log-log plot of the degree distributions 
 for $N=4096$, $\overline{k}=4$ and at various $\alpha$'s.
\label{fig:Pk}}
\end{figure}

To generate the
initial connected network we first add $L_0=N$ links to form a graph
with the topology of a ring (see Fig.~\ref{fig:ex}(a)).
The remaining $L-L_0$ links are added sequentially and randomly, to connect
unlinked nodes. 
The network is evolved by using a local rewiring move (LRM), depicted in 
Fig.~\ref{fig:ex}(f).
     A set of three nodes is randomly selected. First, a node $i$ is selected
     with probability $1/N$, secondly the node $j$ which is a neighbor of $i$
     is selected with probability $1/k_i$ 
     and finally the node $l\neq i$ which is a neighbor of $j$
     is selected with probability $1/(k_j-1)$.
     If $c_{il}=0$, a LRM attempts to delete the $i$-$j$ link of the
     graph ${\cal G}$ and introduce the $i$-$l$ link to obtain the
     new graph ${\cal G}'$ with a probability
\beq 
p = \min \left\{ 1, e^{-\alpha [ H({\mathcal G'}) -H({\mathcal G})] } \right\}
\,,
\label{eq:Metrop}
\eeq 
     where $\alpha$ is a tunable parameter. For $\alpha=0$ the LRM is always
     accepted~\cite{cm03}. 
     In this case the difference between the typical graphs
     and RG's is reflected in the degree distribution. 
     Using a master equation approach \cite{baiesi}, one can show 
     that the degree distribution $P(k)$ indeed decays exponentially, as 
     shown numerically in Fig.~\ref{fig:Pk}. 
     In the opposite limit of $\alpha \to \infty$, LRM 
     strongly favors connecting nodes with degrees as different as
     possible. As a result, usually graphs have several
     high degree nodes (hubs) connected to many other mono-degree nodes
     (leaves). Notice that the LRM cannot split the network in disjoint
     components.

The probability (\ref{eq:Metrop}), the introduction of the Hamiltonian
(\ref{eq:Hamil}) and the LRM have been chosen for their simplicity and
for their analogy with usual rules of equilibrium statistical mechanics, 
but they do not yield an equilibrium distribution of (\ref{eq:Hamil}) at an 
inverse temperature $\alpha$.
However, one can see that the above implementation gives the canonical
distribution of configurations with weights
$\prod_{i=1}^N (k_i-1)! \,e^{- \alpha H(\mathcal G)}$,
hence allowing a description of the system in terms of ensemble of connected
networks at the equilibrium given by the Hamiltonian
$H_{\rm eq}(\mathcal G) \equiv -\alpha^{-1} \sum_{i=1}^N \ln (k_i-1)! + 
H(\mathcal G)$, at temperature $\alpha^{-1}$.
Thus, $H(\mathcal G)$ can be thought as an interaction term from which
we expect the arising of complex correlations in the network.
Due to $H(\mathcal G)$, the fraction of second neighbors
of the node $i$ represented by the first neighbors of node $j$ and of node $l$
(in the LRM, Fig.~\ref{fig:ex}(f)) 
 belongs to the subset of nodes contributing to the energy balance in a LRM.
The non-trivial build up of the correlations necessary to obtain SFN's 
should require $\gamma < 3$, 
meaning an average number of second neighbors $\sim N$.
On the contrary, $\gamma>3$ would forbid the LRM to ``feel'' the global 
structure of the network, hardly giving a fine self-tuning of the network 
correlations. As shown below, indeed we find SFN's with $2<\gamma<3$.

We apply the described dynamical rules to networks composed by
$N$ up to $8192$ nodes and $L=N \overline{k}/2$, 
where the average degree $\overline{k}$ is fixed.
        Thus, we consider sparse networks in the limit of large $N$,
        where the number of links $L$ is much smaller than the maximum
      number $L^{\rm max}_{N} = N(N-1)/2$ of possible links in a $N$-clique.
This is motivated by the case of most of the real networks, 
where typically a link between two nodes is an expensive or rare object.
In our case, we have mainly used $\overline{k}=4$, such that $L=2N$.

        After a large number $\tau \approx 2N^2$ of LRM attempts on the
        initial configurations with $\overline{k} = 4$, we observe a 
        significant reorganization of the entire network structure.
For  $t\gtrsim \tau$ iterations of the LRM, equilibrium is reached, 
as indicated by the 
stable shape of the distribution $P_{\alpha,N}(k)$ of the degrees
(see Fig.~\ref{fig:Pk}).
The most interesting region is  $0.8\lesssim \alpha \lesssim 4.0$, 
where  $P_{\alpha,N}(k)$ appear as power laws. 
By increasing $\alpha$, their slopes decrease and their shape
 changes at large $k$'s, where a shoulder grows, indicating the enhanced 
tendency in the network to form hubs, as expected.
For $\alpha\gtrsim 4.0$ the  fraction of hubs is even more consistent
and the shoulder at high degrees in the $P_{\alpha,N}(k)$ is substituted by a 
bump (see for instance the curve for $\alpha=8$ in Fig.~\ref{fig:Pk}). 
Contrary to other models~\cite{bianconi}, an analysis of the mean degree 
of the largest hub indicates that here
it does not attract  a finite fraction of links, for $N\to \infty$.

\begin{figure}[!b]
\vskip 0.0truecm
\includegraphics[angle=-90,width=\WIDTHB]{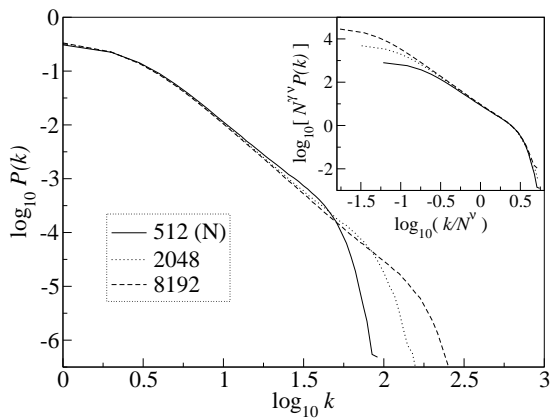}
\caption{Degree distributions for $\alpha=1$. 
Inset: distributions rescaled~(\ref{eq:FSS}) with
the $\gamma$ and $\nu$ values quoted in Table \ref{tab:1}.
\label{fig:Pk2}}
\end{figure}

We now focus on the 
 range $0.8\lesssim \alpha \lesssim 4.0$, where the degree distributions
are broad and power-law like.
To support this SF picture we plot Figure~\ref{fig:Pk2}, in which
a remarkable feature of this model is evident, namely,
the self-similar shape of the $P_{\alpha,N}(k)$ for fixed $\alpha$. 
This is consistent with a finite size form 
\beq
P_{\alpha,N}(k) \simeq k^{-\gamma(\alpha)} f\left(k/N^{\nu(\alpha) } 
\right)\,,
\label{eq:FSS}
\eeq 
where $f(x)$ 
is a scaling function giving a cutoff for sufficiently large values
of $x$, and $f(x)\sim{\textit{const}}$ for $x\to 0$.
In order to quantify the SF nature of the networks
and to support the proposed scaling (\ref{eq:FSS}), we perform a finite size 
analysis, first extrapolating the value of the  distributions slope
at fixed $\alpha$ and for $N\to\infty$.
In Table \ref{tab:1} we show the results.
For each $\alpha$, the value of $\gamma$ so obtained is the starting point 
for attempting a data collapse, by using a rescaling of the form
$N^{\gamma \nu} P(k)$ vs ${k/N^\nu}$.
The values of $\gamma$ that give better collapses (see Table \ref{tab:1})
turn out to be close to the extrapolated values, 
such that the whole picture is consistent.
In addition, we note that the data collapses get worse close to the boundaries
of the SF region $0.8\lesssim\alpha\lesssim4.0$, while
outside this range we can not make good rescalings.  
This supports our first, subjective delimitation of the SF region 
(we stress that the $\overline{k}=4$ case is treated).

\begin{figure}[!t]
\includegraphics[angle=-90,width=\WIDTHA]{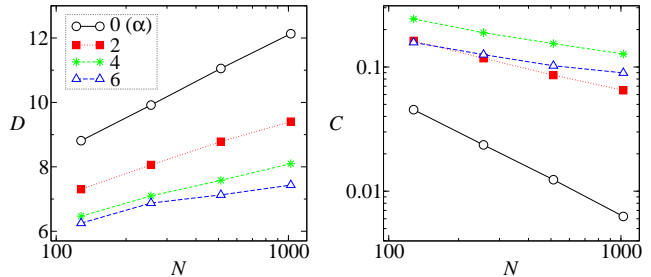}
\caption{
Mean diameter ${\cal D}$ and mean clustering coefficient $C$ 
as a function of $N$
(linear-log and log-log plot, respectively),
for the same set of $\alpha$ values.
\label{fig:C_d}}
\end{figure}
\begin{table}[!b]
\caption{\label{tab:1} 
Numerical evaluation of the exponents of Eq.~(\ref{eq:FSS}).}
\begin{ruledtabular}
\begin{tabular}{c|lllll}
$\alpha$&    0.8&   1&   2&   3&    4\\
\hline
$\gamma$\footnotemark[1]  & 2.9(2)& 2.8(2) & 2.4(1) & 2.3(2) & 2.1(3) \\
\hline
$\gamma$\footnotemark[2] & 3.0 & 2.8 & 2.3 & 2.2 & 2.2\\
$\nu$\footnotemark[2] & 0.4 & 0.45 & 0.55 & 0.58 & 0.58
\end{tabular}
\end{ruledtabular}
\footnotetext[1]{Values extrapolated for $N\to\infty$.}
\footnotetext[2]{Values that give the best data collapse.}
\end{table}

It is interesting to examine the other stylized features of networks.
In Figure \ref{fig:C_d} we plot the mean diameter ${\cal D}(N)$ (the maximum of
the shortest paths between any nodes of a network) 
as a function of $\log N$ for the some representative $\alpha$ values. 
The curves are consistent with a scaling ${\cal D}(N)\sim A+B \log N$ (it seems
to be sub-logarithmic~\cite{cohen} for high $\alpha$). 
Thus, not surprisingly the small world picture is recovered also in our model.

Given a node $i$ connected with $k_i$ neighbors, if $m_i$ is the number of
links between these neighbors, one can quantify the (local) degree
of clustering by the clustering coefficient
$C_i \equiv m_i / L^{\rm max}_{k_i}$.
The mean clustering coefficient $C$ is then the average of the $C_i$
over all nodes of all graph realizations at a given $\alpha$.
In Figure \ref{fig:C_d} we also
 plot $C$ as a function of $N$, in a log-log scale.
The plots are compatible with 
 $C\sim N^{-\sigma(\alpha)}$ for $N\to \infty$, with
$\sigma$ ranging from $\approx 1$ to $\approx0.25$ for $0\leq\alpha\leq6$.
For the studied $N$'s, we notice that  the highest clustering is found
close to $\alpha=4.0$, while it is smaller for high and low $\alpha$.

\begin{figure}[!t]
\includegraphics[angle=-90,width=\WIDTHB]{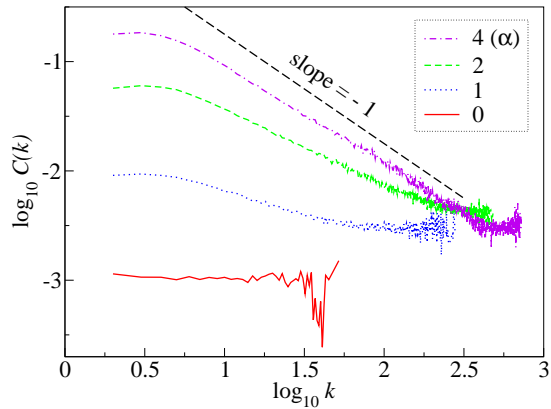}
\caption{Clustering coefficient as a function of the degree, 
for four values of $\alpha$ and $N=8192$.
\label{fig:Ck}}
\end{figure}

Our model also shows the power law dependence
\beq
C(k)\sim k^{-\beta}   \label{eq:Ck}
\eeq 
 indicating that nodes with few links are typically
well clustered while hubs hardly are related to high clustering.
The concept of modularity was introduced to account for the 
hierarchical clustering found in many networks~\cite{ravasz03,szabo}.
In this contexts, networks are built of rather well identifiable clusters,
which are themselves composed by clustered subunits, and so on.
While in Figure \ref{fig:Ck} we see that $\beta=0$ for $\alpha=0$, as for RG,
for increasing $\alpha$ a scaling (\ref{eq:Ck}) takes place in a non negligible
interval of $k$, with a $\beta$ raising with $\alpha$ up to values close
to $\beta=1$. Hence, in a present issue on whether $\beta=1$ is 
universal~\cite{ravasz03}, our result is in favor of the non-universal
character of this exponent.

In summary, we have described a static network model with a dynamics
favoring networks with disassortative mixing and high clustering, 
where at least
three phases can be identified by increasing the parameter $\alpha$,
respectively: exponential, scale free and hub-leaves regimes.
The scale free regime has a signature of modularity  
(the clustering coefficient scales as a power law of the degree with 
a non-trivial exponent),
and the exponent $\gamma$ is comprised in the interesting
range $2<\gamma<3$.
Thus, many of the characteristics displayed by real networks,
usually associated to growing networks with preferential attachment, 
can be obtained as well in networks with a fixed number of nodes,
by using a random rewiring that does not require
preferential attachment.

We gratefully acknowledge useful discussions with A.~L.~Stella.
M.~B.~acknowledges the support of MIUR-COFIN01 and INFM-PAIS02.

\end{document}